\documentclass[twocolumn,preprintnumbers,superscriptaddress,nofootinbib,aps,prd,floatfix]{revtex4}
\pdfoutput=1

\usepackage{amsmath,amssymb}
\usepackage{wasysym}
\usepackage{graphicx}
\usepackage{color,array,subfigure,slashed}
\usepackage[scientific-notation=true]{siunitx}
\usepackage[dvipsnames]{xcolor}
\usepackage{ulem}

\newcommand{\be}{\begin{eqnarray}}
\newcommand{\ee}{\end{eqnarray}}

\newcommand{\bee}{\begin{eqnarray}}
\newcommand{\eee}{\end{eqnarray}}
\newcommand{\beeq}{\begin{equation}}
\newcommand{\eeeq}{\end{equation}}

\newcommand{\beq}{\begin{equation}}
\newcommand{\eeq}{\end{equation}}
\newcommand{\beqn}{\begin{eqnarray}}
\newcommand{\eeqn}{\end{eqnarray}}

\begin{document}

\title{Proton decay testing low energy SUSY}

\begin{abstract}
We show that gauge coupling unification in SUSY models can make
a non-trivial interconnection between collider and proton decay experiments.
Under the assumption of precise gauge coupling unification in the MSSM,  
the low energy SUSY spectrum and the unification scale are intertwined,
and the lower bound on the proton lifetime 
can be translated into {\it upper} bounds on SUSY masses.
We found that the current limit on $\tau(p \to \pi^0　e^+)$
already excludes gluinos and winos $heavier$ than $\sim 120$ and 40 TeV, respectively,
if their mass ratio is $M_3/M_2 \sim 3$.
Next generation nucleon decay experiments are expected to bring these upper bounds 
down to $\sim 10$ and 3 TeV.
\end{abstract}

\author{Stefan Pokorski} 
\affiliation{Institute of Theoretical Physics, Faculty of Physics,\\University of Warsaw, ul.~Pasteura 5, PL--02--093 Warsaw, Poland\\[0.1cm]}
\affiliation{Theoretical Physics Department, CERN, CH-1211 Geneva 23, Switzerland\\[0.1cm]}

\author{Krzysztof Rolbiecki} 
\affiliation{Institute of Theoretical Physics, Faculty of Physics,\\University of Warsaw, ul.~Pasteura 5, PL--02--093 Warsaw, Poland\\[0.1cm]}

\author{Kazuki Sakurai} 
\affiliation{Institute of Theoretical Physics, Faculty of Physics,\\University of Warsaw, ul.~Pasteura 5, PL--02--093 Warsaw, Poland\\[0.1cm]}

\preprint{CERN-TH-2017-154}

\maketitle

Proton decay would be the key evidence for grand unified theories (GUTs) \cite{Langacker:1980js}. 
Among possible decay channels, a special role is played by the $p \rightarrow \pi^0 e^+$ mode for which the dominant contribution may come from the $D=6$ operators depending almost exclusively on the $X,Y$ boson mass and the unified gauge coupling.  
This is in contrast to the other channels induced by $D=5$ operators,
which depend on many more parameters, though the rate is typically larger than
the $p \rightarrow \pi^0 e^+$ mode.  

The main point we want to emphasise and make very explicit in this Letter is that $\tau_{p \rightarrow \pi^0 e^+}$ carries an important information about the low scale supersymmetric (SUSY) spectrum. 
To this end we assume here that the unification of the gauge couplings is 
precise (or exact) within the minimal SUSY Standard Model (MSSM)
without threshold corrections of GUT scale particles \cite{Raby:2009sf}. 
In fact, there exists a class of models where these corrections are absent 
or highly suppressed (see e.g.~\cite{Kawamura:2000ev}).
On the other hand, GUT threshold corrections in conventional models are often too large compared 
to the typical mismatch of gauge couplings at a high scale in the MSSM 
(see \cite{Ellis:2015jwa} for a recent discussion).
This means
that the well-known ``{\it success of gauge coupling unification in the MSSM}'', if not a mere accident,
may favour the aforementioned class of models as 
the correct theory of grand unification.

Under the assumption of precise gauge coupling unification (GCU) in the MSSM, we show that 
the low energy SUSY spectrum and the unification scale are intertwined,
and the lower bound on the proton lifetime $\tau_{p \rightarrow \pi^0 e^+}$
can be translated into {\it upper} bounds on SUSY masses.\footnote{
    Unlike other upper bounds on SUSY masses based on the arguments of 
    the Higgs boson mass \cite{Giudice:2011cg}
    or the neutralino relic abundance \cite{ArkaniHamed:2006mb},
    these bounds dependent neither on the ratio of the Higgs vacuum expectation values,
    $\tan\beta \equiv v_u/v_d$, nor the assumption of $R$-parity conservation and the thermal history of the universe.  
}
This leads to an interesting
interconnection between the proton decay experiments and the collider searches, particularly in view of the future progress on both fronts, in cornering supersymmetric spectrum from above and from below.

At the one-loop the gauge couplings at scale $\tilde \mu$ in the MSSM is given by
\begin{equation}
\frac{2 \pi}{\alpha_i(\tilde \mu)} \,=\, \frac{2 \pi}{\alpha_i(m_Z)} - 
b_i \, \ln \Big( \frac{\tilde \mu}{m_Z} \Big) + 
s_i.
\label{eq:rge}
\end{equation}
where 
$\alpha_1 \equiv \frac{3}{5} \alpha_Y$,
$i = 1,2,3$ represents the gauge group, 
$b_i = (\frac{33}{5}, 1, -3)$ are the one-loop $\beta$-function coefficients for the MSSM and
\beq
s_i \, \equiv \, \sum_\eta b_i^\eta \ln \Big( \frac{m_\eta}{m_Z} \Big)
\label{eq:s}
\eeq
are the threshold corrections of SUSY particles. 
For SUSY particle $\eta$, the mass and its contribution to $b_i$
are given by $m_\eta$ and $b_i^\eta$, respectively.

In the special case where all SUSY particles are mass degenerate at $M_s$,
the threshold correction can be written as $s_i = \delta_i \ln (M_s / m_Z)$
with $\delta_i \equiv (b_i - b_i^{\rm SM})$,
where $b_i^{\rm SM} = (\frac{41}{10}, -\frac{19}{6}, -7)$
are the one-loop $\beta$-function coefficients for the Standard Model (SM).
In this case, exact gauge unification 
$\alpha_1(\tilde \mu) = \alpha_2(\tilde \mu) = \alpha_3(\tilde \mu) \equiv \alpha_G^*$
is achieved by the particular values of $M_s$ and $\tilde \mu$: $M_s^*$, $M_G^{\rm deg *}$,
satisfying 
\beqn
\frac{2 \pi}{\alpha_G^*} - \frac{2 \pi}{\alpha_i(m_Z)} 
+ b_i \, \ln \Big( \frac{M^{\rm deg *}_G}{m_Z} \Big) 
\,=\, \delta_i \, \ln \Big( \frac{M_s^*}{m_Z} \Big)
\label{eq:degene_unif}
\eeqn 
for all $i$.
It should be kept in mind that the quantities 
$M_s^*$, $M_G^{\rm deg *}$ and $\alpha_G^*$ are not variables 
but constants defined as the solution to the above three simultaneous equations.  

Coming back to the general case, let us decompose the vector $s_i$ into three independent vectors as 
\cite{Krippendorf:2013dqa}
\beq
s_i \,=\, \delta_i \, \ln \Big( \frac{T}{m_Z} \Big) + b_i \ln \Omega + C\,.
\label{eq:s}
\eeq
The solution to this set of equations is given by
\beqn
\ln \Big( \frac{T}{m_Z} \Big) &=& v_i s_i / D \\
\ln \Omega  &=& u_i s_i / D\\
C  &=& \epsilon_{ijk} \delta_j b_i s_k / D
\eeqn
where summation is understood for the repeated indices and $\epsilon_{ijk}$
is the antisymmetric tensor and
\beqn
v \,=\, 
\begin{pmatrix}
b_2 - b_3 \\ -b_1+b_3 \\ b_1 - b_2
\end{pmatrix},
~~~
u \,=\, 
\begin{pmatrix}
-\delta_2 + \delta_3 \\ \delta_1 - \delta_3 \\ -\delta_1 + \delta_2
\end{pmatrix}, 
\nonumber \\
D ~=~ 
b_2 \delta_1 - b_3 \delta_1 - b_1 \delta_2 + b_3 \delta_2 + b_1 \delta_3 - b_2 \delta_3\,.
\eeqn
Plugging the concrete values of $b_i$, $\delta_i$ and $b_i^\eta$
into these expressions, one gets  
\beqn
T &=& \Big[ M_3^{-28} M_2^{32}  (\mu^{4}
 m_A)^{3}  X_{T} \Big]^{\frac{1}{19}},
 \label{eq:tsusy}
 \\
\Omega &=& \Big[ M_3^{-100} M_2^{60} (\mu^{4} m_A)^{8} X_{\Omega} \Big]^{\frac{1}{288}}\,,
\label{eq:omega}
\\
C &=& \frac{125}{19} \ln M_3 - \frac{113}{19} \ln M_2 - \frac{40}{19} \ln \mu - \frac{10}{19} \ln m_A \nonumber \\
&+& \sum_{i=1...3} \Big[ 
 \frac{79}{114} \ln m_{\tilde d_{Ri}}
- \frac{10}{19} \ln m_{\tilde l_i}
- \frac{121}{114} \ln m_{\tilde q_i} 
\nonumber \\
&+& \frac{257}{228} \ln m_{\tilde u_{Ri}}
+ \frac{33}{76} \ln m_{\tilde e_{Ri}}
\Big]\,.
\label{eq:C}
\eeqn
with
\beqn
X_{T} &\equiv& \prod_{i=1...3} \Big( \frac{m^3_{\tilde l_i}}{m^3_{\tilde d_{Ri}}} \Big) \Big( \frac{m^7_{\tilde q_i}}{m^2_{\tilde e_{Ri}} m^5_{\tilde u_{Ri}}}  \Big)\,,
\\
X_{\Omega} &\equiv& \prod_{i=1...3} \Big( \frac{m^8_{\tilde l_i}}{m^8_{\tilde d_{Ri}}} \Big)
\Big( \frac{m^6_{\tilde q_i}m_{\tilde e_{Ri}}}{m^7_{\tilde u_{Ri}}} \Big)\,.
\eeqn
The SUSY mass parameters appearing in this Letter should be understood 
as the magnitude of the corresponding parameters
because phase factors do not affect RG running.
In most models, the sfermion contributions to $T$ and $\Omega$ are negligible 
(i.e.~$X_T \sim X_\Omega \sim 1$). 
In particular, 
these contributions vanish if the masses are degenerate within the SU(5) multiplets, 
${\bf \bar 5}_i = (\tilde d_R^c, \tilde l)_i$, ${\bf 10}_i = (\tilde q, \tilde u_R^c, \tilde e_R^c)_i$.
One can explicitly check that for a degenerate spectrum, $\ln\Omega = C =0$. 

To see roles of $T$, $\Omega$ and $C$
in gauge unification, we substitute Eq.\,\eqref{eq:s} into Eq.\,\eqref{eq:rge} and obtain
\beqn
\frac{2 \pi}{\alpha_i(\tilde \mu)}
=
\frac{2 \pi}{\alpha_G^*}
- b_i  \ln \Big( \frac{\tilde \mu }{\Omega M^{\rm deg *}_G} \Big) 
+ \delta_i  \ln \Big( \frac{T}{M_s^*} \Big) 
+ C,
\eeqn
where Eq.\,\eqref{eq:degene_unif} has also been used.
It is clear that the exact unification for the general case is obtained when the right-hand-side (RHS) becomes $i$-independent, that is at $T = M_s^*$ \cite{Carena:1993ag}  and the exact unification scale is given by
\beq
M_G \,=\, \Omega M^{\rm deg *}_G \,.
\label{eq:rel1}
\eeq
The unified gauge coupling is related to that of the degenerate case as
\beq
\alpha^{-1}_G \,=\, \alpha^{*-1}_G +\, \frac{C}{2 \pi}.
\label{eq:rel2}
\eeq

Away from the exact unification, 
we define a candidate unification scale $M_U$ and a semi-unified coupling $\alpha_U$ by
$\alpha_1(M_U)=\alpha_2(M_U) \equiv \alpha_U$.
This scale can be computed from a low energy spectrum as
$M_U = \Omega M^{\rm deg*}_G (T/M_s^*)^{\frac{\delta_1 - \delta_2}{b_1 - b_2}}$,
and at this scale the gauge couplings are given by
\beq
\frac{2 \pi}{\alpha_i(M_U)} \,=\,
\frac{2 \pi}{\alpha_G^*}
\,+
\Big( \delta_i - b_i \frac{\delta_1 - \delta_2}{b_1 - b_2} \Big) \ln \Big( \frac{T}{M_s^*} \Big) +\, C\,.
\eeq
Using this formula, a measure of gauge coupling unification, 
which we define as $\epsilon_3 \equiv (\alpha_3(M_U) - \alpha_U)/\alpha_U$,
is calculated as 
\beqn
\epsilon_3 \,=\, \frac{\alpha_G^*}{2 \pi} \,Y\, \ln \Big( T/M_s^* \Big) 
\,+\, \cdots,
\label{eq:e3_rep}
\eeqn
where the dots represent higher order terms of $\frac{\alpha_G^*}{2 \pi}$ and 
\beq
Y ~\equiv~ \frac{b_1 (\delta_2 - \delta_3) + b_2 (-\delta_1 + \delta_3) + b_3 (\delta_1 - \delta_2)}{b_1 - b_2}\,.
\eeq
It is interesting that $\epsilon_3$ depends only on $T$
at the leading order \cite{Carena:1993ag}.

Our argument so far is based on the one-loop renormalization group equations (RGEs).
It turns out that the relations Eqs.\,\eqref{eq:rel1}, \eqref{eq:rel2} and \eqref{eq:e3_rep}
still hold numerically with a good accuracy at two-loop level  
if the constants are replaced by the two-loop corrected values:
$M_s^* = 2.08$\,TeV,
$M_U^{\rm deg *} = 1.27 \cdot 10^{16}$\,GeV and
$\alpha_G^{*-1} = 25.5$.
\begin{figure}[t!]
\begin{center}
\includegraphics[scale=0.38]{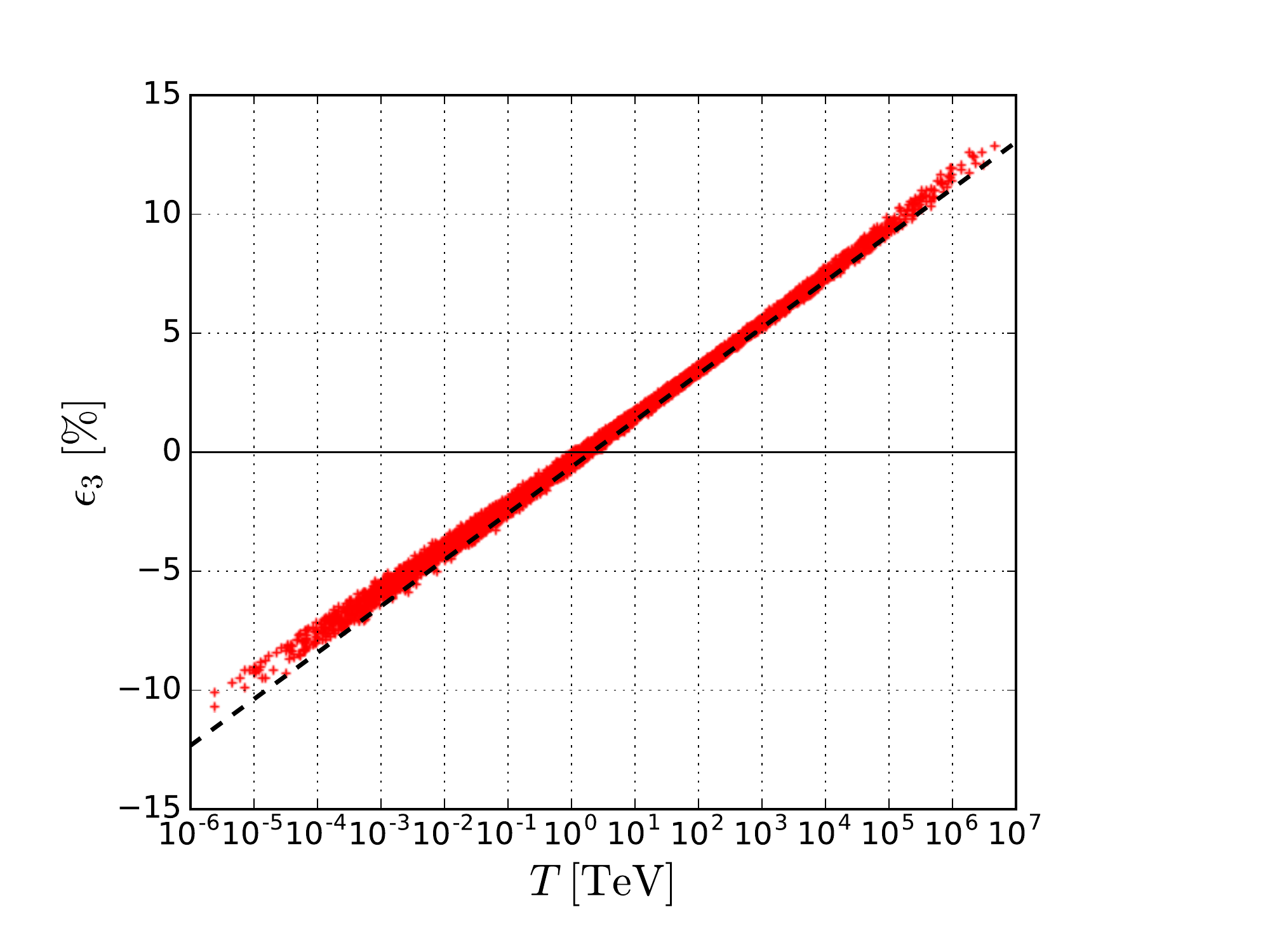} 
\includegraphics[scale=0.38]{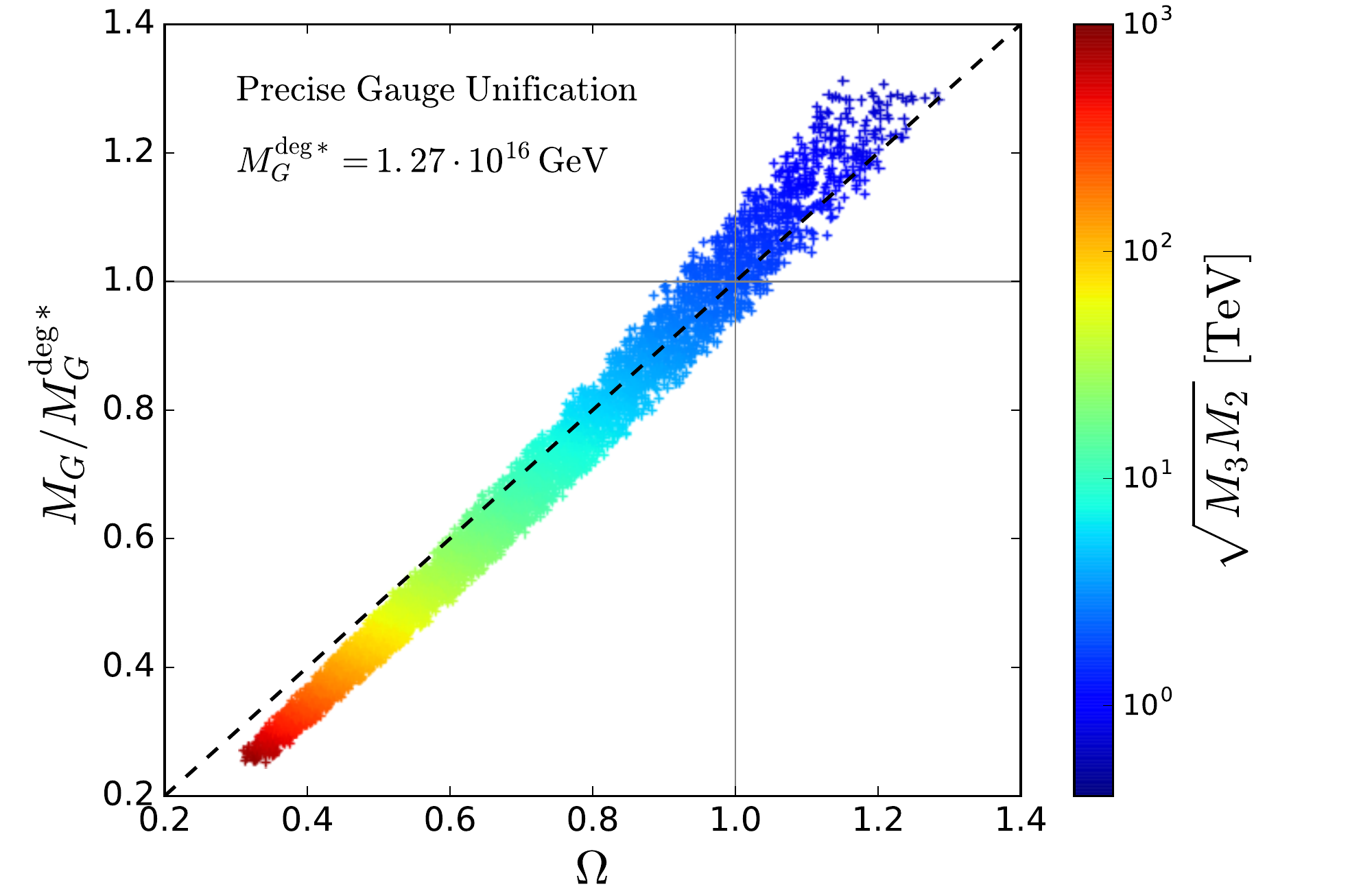} 
\includegraphics[scale=0.38]{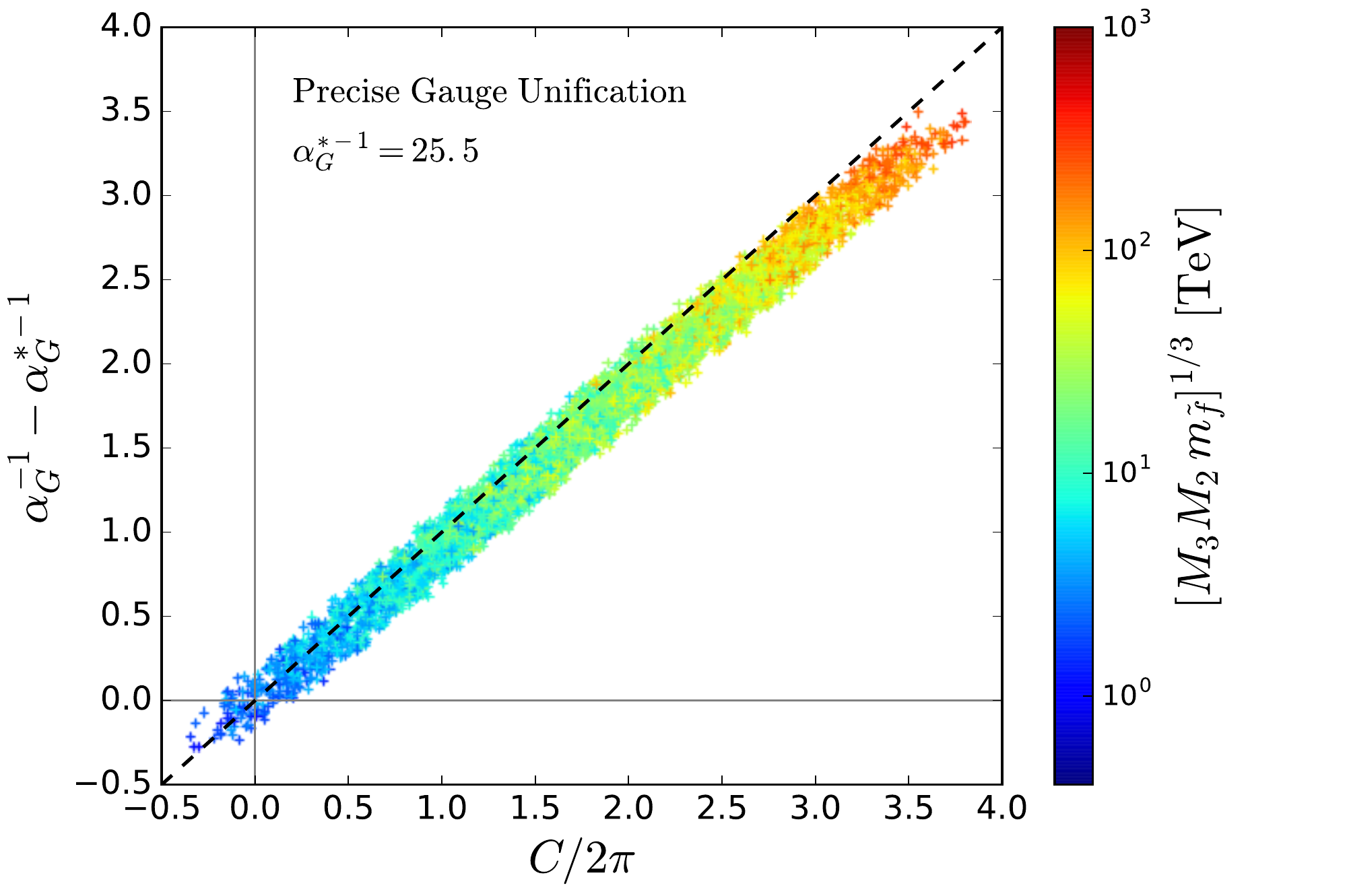} 
\caption{\label{fig:tsusy} \small
A scan of SUSY particle masses projected onto 
the ($T$, $\epsilon_3$) (top),
the ($\Omega$, $M_G/M_G^{\rm deg*}$) (middle)
and ($C/2 \pi$, $\alpha_G^{-1}-\alpha_G^{*-1}$) (bottom) planes.
In the middle and bottom plots, the precise gauge unification $(|\epsilon_3| < 0.1 \%)$
is required, and the colour-codes represent typical SUSY scales 
$\sqrt{M_3 M_2}$ and $(M_3 M_2 m_{\tilde f})^{\frac{1}{3}}$, respectively. 
The dashed lines represent the one-loop relations 
Eq.\,\eqref{eq:e3_rep}, \eqref{eq:rel1} and \eqref{eq:rel2}
for the top, middle and bottom plots, respectively.
}
\end{center}
\end{figure}
We show in Fig.~\ref{fig:tsusy}
the result of our numerical scan.
All numerical scans presented in this Letter use
a two-loop RGE code including the effect of the top Yukawa coupling, following \cite{Langacker:1992rq}.
We use $\tan\beta=10$ but a variation of $\tan\beta$ results in negligible effects.  
The SUSY breaking parameters are uniformly scanned in the logarithmic scale within [$m_{\min}$, $10^3$\,TeV].
We take $m_{\rm min} = 1.5$\,TeV for $M_3$ and 200\,GeV for $M_2, \mu$ and $m_A$. 
The sfermion masses are assumed to be universal ($\equiv m_{\tilde f}$) for simplicity and $m_{\min}=1$\,TeV is used.
We also vary $\alpha_3(m_Z) = 0.1184(7)$, according to 
the 1-$\sigma$ uncertainty.

The top plot in Fig.\,\ref{fig:tsusy}
tests the predicted relation Eq.\,\eqref{eq:e3_rep} (dashed line).
We see that the exact unification occurs 
only when the SUSY masses are arranged such that
$T$ computed by Eq.\,\eqref{eq:tsusy}
is within a certain range [1, 4]\,TeV centred around $\sim 2$\,TeV.
The width of $T$ for exact unification comes mainly from the uncertainty on $M_s^*$
due to the variation of $\alpha_3(m_Z)$.\footnote{
    The constants $M_s^*$, $M_G^{\rm deg*}$ and $\alpha^*_G$ 
    should be understood as functions of $\alpha_3(m_Z)$ in the scan.
}
The middle plot shows the correlation between $\Omega$ and 
the exact unification scale, $M_G$.
Hereafter, we require a precise gauge unification, $|\epsilon_3| < 0.1 \%$.
The dashed line corresponds to Eq.\,\eqref{eq:rel1}.
The colour of points represents a typical SUSY scale $\sqrt{M_3 M_2}$.
One can see that heavy SUSY tends to have a small unification scale.
For the PeV scale SUSY with $\sqrt{M_3 M_2} \sim 10^3$\,TeV, $M_G$ is reduced by a factor of 5
compared to the TeV scale one.
The bottom plot confirms the predicted relation Eq.\,\eqref{eq:rel2} (dashed line).
The colour-code indicates a SUSY scale, $(M_3 M_2 \, m_{\tilde f})^{\frac{1}{3}}$.
We see that high scale SUSY tends to predict a smaller unified coupling, $\alpha_G$, 
but the variation is small and only up to $\sim 10 \%$ between the TeV and PeV scale SUSY mass points. 

An interesting observation follows from 
the last two plots of Fig.\,\ref{fig:tsusy}. 
High scale SUSY, where the unification scale is lower, in general leads to a rapid proton decay, $p \to e^+ \pi^0$. 
This is because the rate $\Gamma(p \to e^+ \pi^0)$ scales as $\alpha_G / (M_G)^4$, 
where the $X,Y$ boson mass is identified as the unification scale, since the precise gauge unification
implies all GUT particles charged under the SM gauge group have the same mass, $M_G$.
Turning this around, the lower limit on $M_G$ from the proton lifetime measurement 
(if found, bearing in mind that the variation of $\alpha_G$ is small)
can place {\it upper bounds} on the masses of SUSY particles. 
Let us denote this lower limit by $M_{PD}$: $M_G > M_{PD}$. 
Then, eliminating $M_3$ from Eq.\,\eqref{eq:omega} by using Eq.\,\eqref{eq:tsusy},
Eq.\,\eqref{eq:rel1} gives us
\beq
M_2^{\frac{4}{5}} (\mu^4 m_A)^{\frac{1}{25}} \, < \,
M_s^* \cdot \Big( \frac{M^{\rm deg *}_G}{M_{PD}} \Big)^{\frac{2016}{475}}
\cdot X_{\rm EW}^{\frac{1}{25}}\,,
\label{eq:bound}
\eeq
where
\beq
X_{\rm EW} ~\equiv\, \prod_{i=1...3} \Big( \frac{m_{\tilde l_i}}{m_{\tilde d_{Ri}}} \Big)
\Big( \frac{m^7_{\tilde q_i}}{m^3_{\tilde e_{Ri}} m^4_{\tilde u_{Ri}}}
\Big) \,.
\eeq
This implies that the smallest mass in the LHS is bounded from above by the RHS of Eq.\,\eqref{eq:bound}.
When this bound is saturated, $M_2 = \mu = m_A$.
The upper limit on the individual parameters are obtained, for example, as
\beqn
M_2 \, < \, M_s^* \cdot \Big(\, \frac{M_s^{*5}}{\mu^{4} m_A } \,\Big)^{\frac{1}{20}} \cdot 
\Big( \frac{M_G^{\rm deg *}}{M_{PD}} \Big)^{\frac{504}{95}}
\cdot
X_{\rm EW}^{\frac{1}{20}} \,.
\eeqn
In this expression the RHS is bounded from above by the experimental lower limit on $\mu$ and $m_A$.

\begin{figure}[t!]
\begin{center}
\includegraphics[scale=0.45]{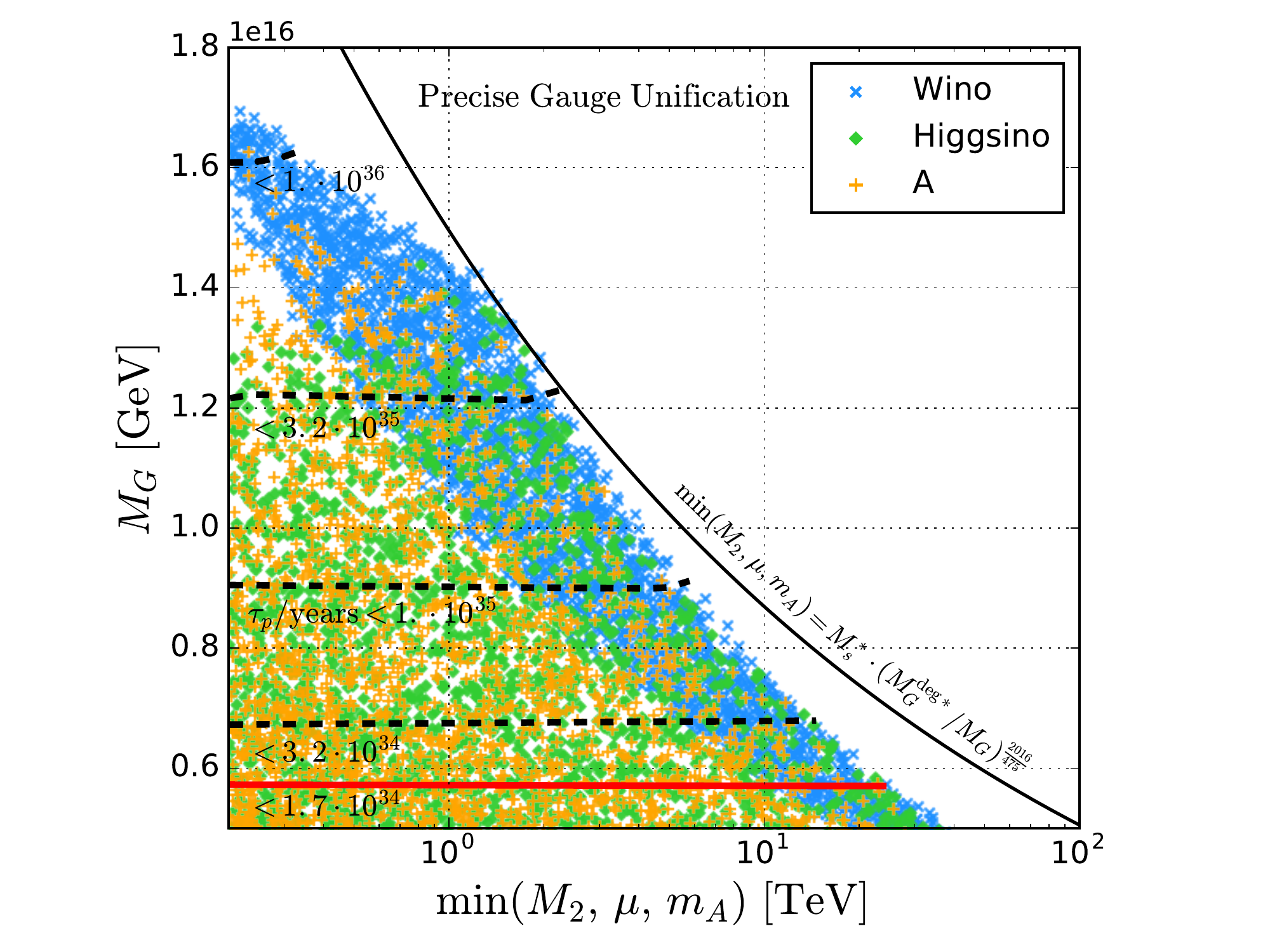} 
\caption{\label{fig:ewmin} \small
Points with precise gauge unification projected onto 
the ${\rm min}(M_2, \mu, m_A)$ vs $M_G$ plane.
The blue, green and orange points correspond to the points where
$M_2$, $\mu$ and $m_A$ is the smallest among them, respectively.
The regions below a black-dashed or a red-solid line are excluded by the
quoted future or current limits on the proton lifetime. 
The black-solid line represents the upper bound found at the one-loop level in Eq.\,\eqref{eq:bound}.
}
\end{center}
\end{figure}

The upper bound Eq.\,\eqref{eq:bound} is observed 
in our numerical scan shown in Fig.\,\ref{fig:ewmin},
where the smallest of $M_2$, $\mu$ and $m_A$
is plotted in the $x$-axis.
The blue, green and orange points correspond to the
cases where $M_2$, $\mu$ and $m_A$ is the lightest among the three,
respectively.
A tendency is observed that
$M_2$ is close to the upper limit if $M_2$ is the lightest.
This is due to the higher power for $M_2$ in Eq.\,\eqref{eq:bound} than for $\mu$ and $m_A$.
At each point we calculate 
$\tau_{p \to \pi^0 e^+}$ based on \cite{Babu:2013jba, Bajc:2016qcc}\footnote{
    The calculation of $\tau_{p \to \pi^0 e^+}$ is not completely model independent.    
    For example, $\tau_{p \to \pi^0 e^+}$ in flipped SU(5) models 
    is smaller by $1/[1 + (1+V_{ud})^2] \sim 1/5$ than in conventional models 
    \cite{Murayama:2001ur}, on which our calculation is based.  
}
using $\alpha_G$ and $M_G$ obtained by the two-loop RGE code.
The horizontal black-dashed and red-solid lines represent the 
boundaries where all points below them
have the lifetime shorter than the quoted values.
In particular, the region below the red line is excluded by
the current limit: $\tau_{p \to \pi^0 e^+} > 1.7 \cdot 10^{34}$ years \cite{Takhistov:2016eqm}.

The upper bound on the gluino mass can be found by eliminating $\mu^4 m_A$ in Eq.\,\eqref{eq:omega} by using Eq.\,\eqref{eq:tsusy} as
\beq
M_3 \, < \, M_s^* \cdot \frac{M_s^*}{M_2} \cdot
\Big( \frac{M_G^{\rm deg *}}{M_{PD}} \Big)^{\frac{216}{19}}
\cdot
X_{\tilde g}^{\frac{1}{4}} 
\label{eq:m3lim}
\eeq
with
\beq
X_{\tilde g} ~\equiv\, \prod_{i=1...3} 
\Big( \frac{m_{\tilde u_{Ri}} m_{\tilde e_{Ri}}}{ m^{2}_{\tilde q_{i}}} \Big)
 \,.
\eeq
As previously, the RHS of Eq.\,\eqref{eq:m3lim} is bounded from above by the experimental lower limit on the wino mass.

\begin{figure}[t!]
\begin{center}
\includegraphics[scale=0.45]{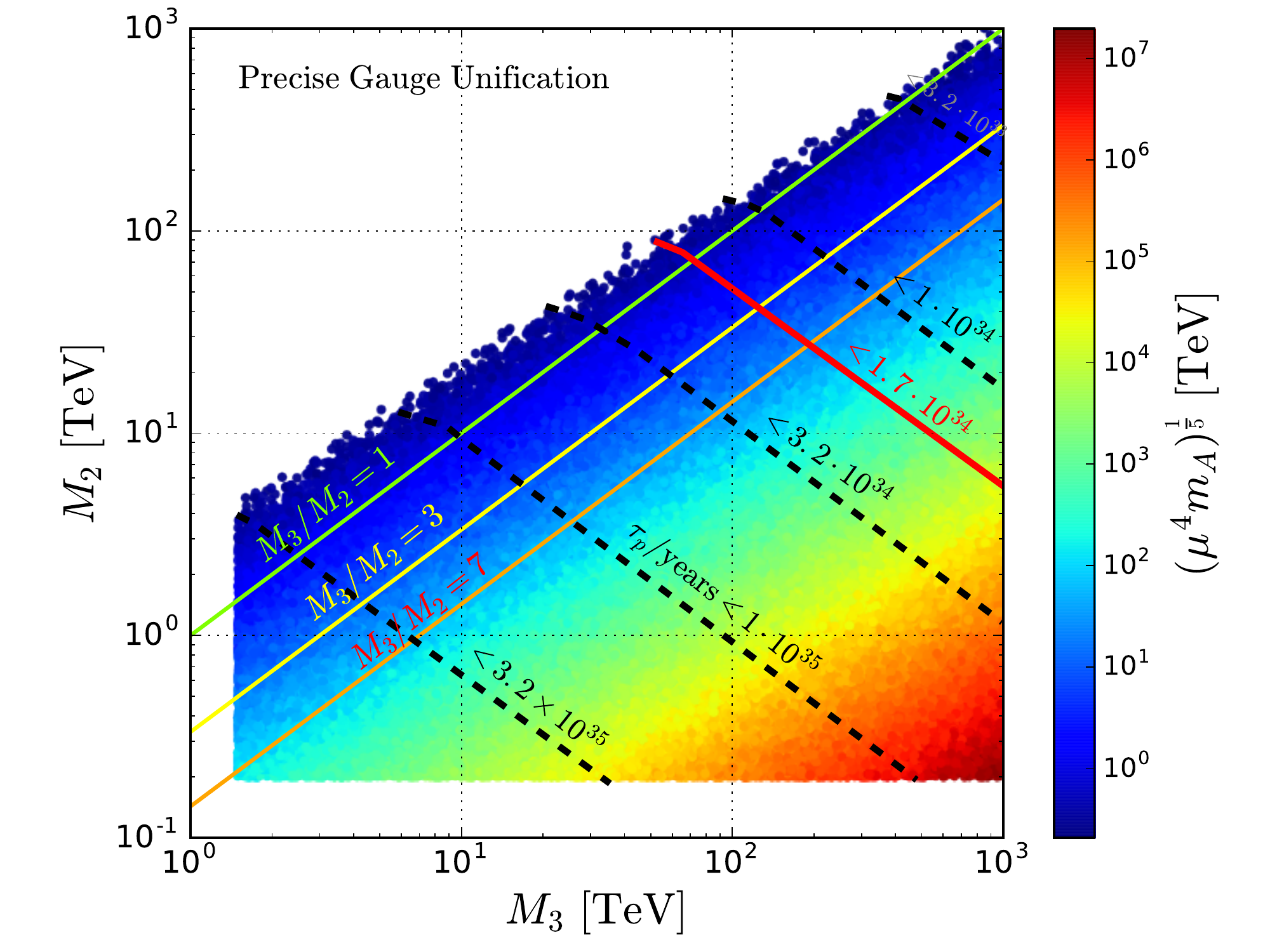} 
\caption{\label{fig:m3-m2} \small
Points with precise gauge unification projected onto the ($M_3$, $M_2$) plane.
The colour-code shows $(\mu^4 m_A)^{\frac{1}{5}}$.
The regions above the black-dashed and red-solid lines are excluded 
by the quoted future or current limits on $\tau_{p \to \pi^0 e^+}$.
The three diagonal lines correspond to $R \equiv M_3/M_2 = 1$, 3 and 7
from top to bottom.  
In this plot the upper boundaries of $\mu$ and $m_A$ scans are extended up to $10^7$\,TeV. 
}
\end{center}
\end{figure}

If the SUSY breaking mechanism is specified, the ratio of gluino and wino masses is usually predicted.
Assuming the value of $R \equiv M_3 / M_2$, the following upper bounds can be derived:
\beqn
M_3 & < & M_s^* \cdot R^{\frac{1}{2}} \cdot
\Big( \frac{M_G^{\rm deg *}}{M_{PD}} \Big)^{\frac{108}{19}}
\cdot
X_{\tilde g}^{\frac{1}{8}} \,,
\\
(\mu^4 m_A)^\frac{1}{5} & < & M_s^* \cdot R^{2} \cdot
\Big( \frac{M_{PD}}{M^{\rm deg *}_G} \Big)^{\frac{144}{90}}
\cdot
X_{\mu}^{\frac{1}{10}} \,,
\label{eq:muma}
\eeqn
where
\beq
X_{\mu} ~\equiv\, \prod_{i=1...3} \Big( \frac{m^2_{\tilde l_i}}{m^2_{\tilde d_{Ri}}} \Big)
\Big( \frac{m^4_{\tilde q_i}}{m_{\tilde e_{Ri}} m^3_{\tilde u_{Ri}}}
\Big) \,.
\eeq

We show in Fig.\,\ref{fig:m3-m2} our scan in the $(M_3, M_2)$ plane with
the colour-code indicating $(\mu^4 m_A)^{\frac{1}{5}}$.
As previously, the black-dashed and red-solid lines represent 
the future and current bounds on $\tau_{p \to \pi^0 e^+}$.
It is evident that $M_3$ and $M_2$ are highly sensitive to the proton lifetime
and constrained by it from above.
This is in direct contrast to collider searches, constraining these parameters 
from below.
Unlike $M_3$ and $M_2$, $\mu$ and $m_A$
are almost insensitive to the proton lifetime, which follows from the lower power of $M_{PD}$
in Eq.\,\eqref{eq:muma}.  
On the other hand, they are highly sensitive to $R$.
In particular, $\mu$ is typically a TeV for $R=1$ 
whereas it is ${\cal O}(100)$\,TeV for $R=7$.
The implication of this to naturalness and phenomenology are studied in detail in 
\cite{Raby:2009sf,Krippendorf:2013dqa,us}.

It is remarkable that the current proton lifetime limit already excludes 
the gluino and wino masses {\it larger} than 200 and 30\,TeV for $R \sim 7$ (e.g.~AMSB)
and 120 and 40\,TeV for $R \sim 3$ (e.g.~CMSSM, GMSB), respectively.  
Next generation nucleon decay experiments are expected to improve the current $\tau_{p \to \pi^0 e^+}$ limit by a factor of ten \cite{Babu:2013jba},
which will result in tightening the upper bounds on gluino and wino masses further down to
$(M_3, M_2) \lesssim (10,\, 3)$\,TeV for $R \sim 3$
and $(M_3, M_2) \lesssim (15,\, 2)$\,TeV for $R \sim 7$.
These bounds are close to the lower mass limits
$(M_3, M_2) \gtrsim (10,\, 2.7)$\,TeV \cite{Cohen:2013xda,Low:2014cba},
which are expected to be obtained at future 100\,TeV hadron-hadron colliders. 

We have investigated the link between the proton lifetime $\tau_{p \to \pi^0 e^+}$ and the supersymmetric spectrum under the assumption of vanishing GUT thresholds. 
It has been shown that most of the allowed mass range of gluinos and winos will be probed by future collider and proton lifetime experiments.
It will also be interesting to extend this study to models with non-vanishing GUT threshold corrections 
(see e.g.~\cite{Hisano:2013cqa}).


\section*{Acknowledgments}
\vspace{-4mm}
KS thanks Zackaria Chacko, Kiwoon Choi, Sebastian Ellis, Shigeki Matsumoto and James Wells
for helpful discussion.
The work of SP and KS is partially supported by the National Science Centre, Poland, under research grants
DEC-2014/15/B/ST2/02157 and DEC-2015/18/M/ST2/00054.
The work of KR and KS 
is supported by the National Science Centre (Poland) under Grant 2015/19/D/ST2/03136.



\end{document}